\documentclass{ws-procs9x6}

\begin{document}

\title{Shock acceleration in partially neutral plasmas}

\author{G. MORLINO$^*$, E. AMATO, P. BLASI and D. CAPRIOLI}

\address{INAF, Osservatorio Astrofisico di Arcetri,\\
L.go E. Fermi-5, Firenze, 50125, Italy\\
$^*$E-mail: morlino@arcetri.astro.it}

\begin{abstract}
We present the non-linear theory of shock acceleration applied to SNRs expanding
into partially neutral plasma. Using this theory we show how the Balmer lines
detected from young SNRs can be used to test the efficiency of shocks in the
production of cosmic rays. In particular we investigate the effect of
charge-exchange between protons and neutral hydrogen occurring in the
precursor formed ahead of the shock. In this precursor the CR pressure
accelerate the ionized component of the plasma and a relative velocity between
protons and neutral hydrogen is established. On the other hand the
charge-exchange process tends to equilibrate ions and neutrals resulting in the
heating of both components. We show that even when the shock converts only a few
per cent of the total bulk kinetic energy into CRs, the heating is efficient
enough to produce a detectable broadening of the narrow Balmer lines emitted by
the neutral hydrogen.
\end{abstract}


\bodymatter

\section{Introduction}\label{sec:intro}

Supernova remnants (SNRs) are thought to be the primary sources of Galactic
cosmic rays (CRs). Diffusive shock acceleration (DSA) occurring around the
forward shock is considered the most promising  mechanism for the
acceleration of particles up to very high energies. In order to explain the flux
of CRs observed at the Earth, the shock should convert a fraction $\sim 10-30\%$
of the plasm kinetic energy into non thermal particles. Indeed, from the
theoretical point of view, DSA can be very efficient in producing
accelerated particles, but a clear confirmation of this prediction still lacks,
even if many circumstantial evidences have been collected (see e.g. Ref.
\refcite{morlino09}).
A possible way to investigate the efficiency of CRs production is through the
analysis of the Balmer lines associated with shocks of SNRs which propagate in a
partially ionized plasma. 
The hydrogen lines of a Balmer shock consist of two superimposed components:
the narrow component is emitted by neutral hydrogen after entering the shock
front and the broad component by hot protons after undergoing charge exchange
with incoming neutral hydrogen atoms. The width of the broad component reflects
the proton temperature behind the shock front, while the width of the narrow
component provides a direct measurement of the temperature upstream of the shock
\cite{Chevalier80, Heng09}. 

We stress that, when the shock is not modified by the presence of CRs, the
charge exchange process and the Balmer emission both occur only downstream of
the shock. On the other hand, when CRs are accelerated efficiently, the shock
structure is modified and a precursor is generated upstream: the CRs
pressure slows down and compresses the ionized plasma. Even if the neutral
component is not directly affected by the CR pressure, a relative velocity
between ions and neutrals is now established. Hence charge exchange can
also occur upstream of the shock, resulting in the pre-heating of neutral
hydrogen. We do expect two remarkable consequences: 1) the narrow Balmer lines
can be emitted also from the upstream region and 2) the typical width of these
lines becomes larger with respect to the case without CRs, because the hydrogen
temperature increases.

A further consequence of efficient acceleration is that the temperature of
the shocked ions downstream is lower with respect to the case with no
acceleration. This occurs because a non negligible fraction of the shock
kinetic energy is converted into CRs, rather than into thermal energy. As a
consequence also the width of the broad Balmer lines is affected, being reduced
with respect to the case with no acceleration.

Remarkably all these aspects have been observed in some Balmer-dominated
shocks:1) from a region of the Tycho remnant, the emission of narrow Balmer
lines has been detected upstream of the shock \cite{Lee10}; 2) in several SNRs
the narrow Balmer lines present a width incompatible with the typical
temperature of the interstellar medium \cite{Sollerman03}; 3) in two different
cases where the Balmer emission has been detected, i.e. RCW 86 \cite{Helder09}
and SNR 0509-67.5 \cite{Helder10}, the width of the broad lines led to a
downstream temperature which is lower than that estimated form the measurement
of the shock proper motion, suggesting that a fraction of the total energy is
converted into CRs.

In order to quantify all these effects simultaneously, in this paper we present
the solution for the stationary case of DSA when the shock propagates into a
partially ionized plasma. We use a semi-analytical method, similar to those used
in Ref.~\refcite{Amato06}, specialized for plane shock geometry.

\section{Interaction between ions and neutrals in the precursor}\label{sec:CE}
The process of charge exchange can occur in the precursor only if its typical
length scale is smaller than the precursor length. The length of the precursor
created by CR pressure is of the order of the diffusion length of the particles
with maximum energy, i.e.:
\begin{equation} \label{eq:L_prec}
 L_{prec} \simeq D(E_{\max})/u_0 = 
 3 \cdot 10^{17} E_{\rm TeV} B_{\mu G}^{-1} u_8^{-1} \, {\rm cm}\,,
\end{equation}
where the maximum energy is expressed in TeV, the magnetic field in $\mu$G and
the shock speed is $u_0= u_8 10^{8}\rm cm/s$. We assume that particles diffuse
with a Bohm-like diffusion coefficient, $D(p)= (c/3) r_L(p)$, where $r_L= pc/eB$
is the Larmor radius and the magnetic field strength is assumed constant in the
precursor. Let call $n_H$ the neutral hydrogen density, the length scale for a
proton to make charge exchange upstream is:
\begin{equation} \label{eq:L_ce}
 L_{ce} = u_0/\left( n_H\, \sigma_{ce}(v_r) \, v_r\right)
        \simeq 10^{15} (u_0/v_r) (n_H/1{\rm cm^{-3}})^{-1} \, {\rm cm}\,.
\end{equation}
We note that when the relative speed between ions and neutral hydrogen is
$v_r\lesssim 2000$ km/s, the charge exchange cross section is roughly
constant \cite{Heng09}. Comparing $L_{ce}$ and $L_{prec}$ we see that a single
proton can charge exchange several times in the precursor before crossing the
shock. In this work we neglect the role of ionization because the ionization
cross section is much smaller than the charge exchange length when $v_r\lesssim
1000$ km/s, but we will defer a more extensive treatment of ionization to a
future paper.

\section{Hydrodynamics with charge-exchange}  \label{sec:fluid}
The behaviour of a system of neutral hydrogen and protons interacting via charge 
exchange is described by a kinetic equations for each species:
\begin{equation} \label{eq:kin}
 \frac{\partial f_i}{\partial t} + v_x \,  \frac{\partial f_i}{\partial x}=
 \beta_i(x,{\bf v}) f_j(x,{\bf v}) - \beta_j(x,{\bf v}) f_i(x,{\bf v})\,,
\end{equation}
where $i,j= H, p$ and $f_i$ is the distribution function. Eq.~(\ref{eq:kin}) is
written in one spatial dimension because we are considering plane shocks
propagating along the $x$ direction. As a consequence $f_i$ is only a function
of the position $x$ and of the particle velocity. The charge exchange frequency
is defined by:
\begin{equation} \label{eq:beta}
 \beta_i(x, {\bf v}) =  \int d{\bf w} \,|{\bf v} - {\bf w}| 
       \sigma_{ce}(|{\bf v}-{\bf w}|) f_i(x,{\bf w})\,.
\end{equation}
Notice that the charge exchange cross section is only a function of the relative
speed $v_r \equiv |{\bf v}_H-{\bf v}_p|$. Now we consider only the stationary
solution, hence $\partial f_i/\partial t=0$. From the integration of first,
second and third moments of Eq.~(\ref{eq:kin}) we get the fluid equations for
neutral hydrogen:
\begin{eqnarray}
 \frac{\partial}{\partial x} \left[ \rho_H u_H \right] = 0\,,
                                                         \label{eq:FE_H1} \\
 \frac{\partial}{\partial x} \left[ \rho_H u_H^2 + P_H \right] = -q_m \,,
                                                         \label{eq:FE_H2} \\
 \frac{\partial}{\partial x} 
    \left[ \frac{1}{2} \rho_H u_H^3 + \frac{\gamma}{\gamma-1} P_H u_H \right]
    = -q_e  \,.                                          \label{eq:FE_H3}
\end{eqnarray}
The term $q_m$ and $q_e$ are the flux of momentum and energy which are
transferred from neutral hydrogen to protons, i.e: 
\begin{eqnarray} \label{eq:qs}
 q_m = \int dv_H dv_p \, (v_H-v_p) \, \sigma_{ce}(v_r) v_r f_H(v_H) f_p(v_p) \\
 q_e = \int dv_H dv_p \, \frac{1}{2}(v_H^2-v_p^2) \, \sigma_{ce}(v_r) v_r 
       f_H(v_H) f_p(v_p) \,.
\end{eqnarray}
In order to get an analytic expression for $q_m$ and $q_e$ we follow
Ref.~\refcite{Pauls95}. First of all we assume that both hydrogen and
protons distributions are Maxwellian, with a bulk velocity ${\bf u}_i(x)$,
along the $x$ direction, and a temperature $T_i(x)$, i.e. 
  $ f_i(x,{\bf v}) = \rho_i \,(\pi \, v_{T_i}^2)^{-3/2} 
                   e^{ -({\bf v} - {\bf u}_i)^2/{v_{T_i}^2} }$,
where the thermal speed is defined as $v_{T_i}=\sqrt{2P_i/\rho_i}$. Moreover we
note that for $v_r\lesssim 2000$ km/s, the charge exchange cross section
decreases only logarithmically with increasing $v_r$, hence we can approximate
$\sigma_{ce}$ as a constant evaluated at the average relative speed and pull it
out of the integrals. Under this assumptions momentum and energy transfer can be
expressed as follows:
\begin{eqnarray}
 q_m = \sigma_{ce}(\bar u)\rho_p \rho_H (u_H-u_p) \left[
       \bar u + v_{T_H}/h_1(H,p) + v_{T_p}/h_1(H,p) \right]  \,; \\
 q_e = \sigma_{ce}(\bar u)\rho_p \rho_H  \frac{1}{2} 
      \left\{
       \bar u (u_H^2-u_p^2) +\frac{3}{2} 
       \left[ v_{T_H}^2 h_2(p,H) - v_{T_p}^2 h_2(H,p) \right] +
                                                        \right. \nonumber \\
       \left. + 2(u_H-u_p) \left[ \frac{u_H v_{T_H}^2}{h_1(p,H)} 
         +\frac{u_p v_{T_p}^2}{h_1(H,p)} \right] 
      \right\}
\end{eqnarray}
where the average relative speed $\bar u$, and the functions $h_{1,2}$ are
defined as:
\begin{eqnarray}
 \bar u  = \left[ \frac{4}{\pi} (v_{T_i}^2+v_{T_j}^2) + (u_i-u_j)^2 
           \right]^{1/2}   \,; \\
 h_1(i,j)= \left[ 4 \left(\frac{4}{\pi} v_{T_i}^2 + (u_i-u_j)^2 \right)+ 
            \frac{9\pi}{4} v_{T_n}^2           
           \right]^{1/2} \,; \\
 h_2(i,j)= \left[ \frac{4}{\pi} v_{T_i}^2 + \frac{64}{9\pi} v_{T_j}^2 
            + (u_i-u_j)^2 
           \right]^{1/2} \,. 
\end{eqnarray}

The fluid equations for protons are similar to those for hydrogen, but include
the contribution due to CRs. Following Ref.~\refcite{Amato06} it is easy to show
that the flux conservation of mass, momentum and energy are:
\begin{eqnarray}  
 \frac{\partial}{\partial x} \left[ \rho_p u_p \right] = 0  \label{eq:FE_p1} \\
 \frac{\partial}{\partial x} \left[ \rho_p u_p^2 + P_p + P_c \right] 
    = q_m                                                   \label{eq:FE_p2} \\
 \frac{\partial}{\partial x} \left[ 
    \frac{1}{2} \rho_p u_p^3 + \frac{\gamma}{\gamma-1} P_p u_p \right]
    = - u_p \frac{\partial P_c}{\partial x} + q_e  \,.      \label{eq:FE_p3}
\end{eqnarray}
where $P_c$ is the CR pressure which is connected to the CR distribution
function through $P_c(x)= 4\pi/3 \int p^2dp\, pv f_c(x,p)$. Equations
(\ref{eq:FE_H1})-(\ref{eq:FE_H3}) and (\ref{eq:FE_p1})-(\ref{eq:FE_p3}) are
coupled through $q_m$ and $q_e$ and can be solved numerically to get the
pressures and the bulk velocities of both protons and hydrogen as implicit
functions of $P_c$. In order to close the system of equations we need to solve
also the transport equation for the CR distribution function, $f_c$. All the
system can be solved using iterative techniques similar to those used in
Ref.~\refcite{Amato06}. We will describe the full procedure in a more detailed
work.

It is worth stressing that the technique used allows one to take into account
other important effects in the shock acceleration theory, like magnetic
field amplification and turbulent heating. For the sake of simplicity here
we neglect such complications in order to isolate the role of charge
exchange.

\section{Results and Conclusions}\label{sec:disc}
\begin{figure}
\begin{center}
\psfig{file=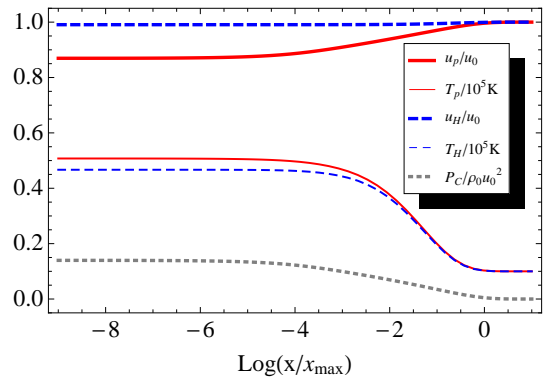,width=3.2in}
\caption{Profiles of velocities and temperatures of hydrogen and protons in the
precursor upstream of the shock. The thick-dotted line shows the CR pressure.
All quantities are normalized as shown in the legend. The typical
precursor length is $x_{\max}= D(p_{\max})/u_0$.}
\label{fig:1}
\end{center}
\end{figure}

We consider a typical case of shock with the following parameters: shock
speed $u_0=2000$ km/s, total upstream density $\rho_0= 1 \rm cm^{-3}$, neutral
fraction 50\%, upstream temperature $T_0=10^4$ K, upstream magnetic field
strength $B_1= 10 \mu$G and maximum momentum of accelerated protons fixed to
$p_{\max}=10^4 m_pc$. The parameter which regulates the injection efficiency is
taken $\xi_{\rm inj}= 3.95$ \cite{Amato06}. 
For these values we get $L_{prec}/L_{ce}\simeq 150$, hence we do expect an
efficient role of charge exchange, while $L_{prec}/L_{ion}\simeq 10^{-4}$
hence the ionization can be neglected as stated above. 
In Fig.~\ref{fig:1} we show the velocity and the temperature profiles in the
precursor for both protons and hydrogen. The dotted line shows the normalized CR
pressure, which at the sub-shock position reaches the value $P_c/\rho_0
u_0^2=0.14$, hence we are dealing with a mildly efficient shock.
The upstream temperatures of protons and hydrogen reach the values of $5 \cdot
10^5$ K and $4.7 \cdot 10^5$ K, respectively. Protons are slightly wormer than
hydrogen because they also suffer the adiabatic compression due to the CR
pressure. The narrow Balmer line width corresponding to the hydrogen temperature
at ths subshock is 46 km/s, to be compared with 21 km/s resulting in the absence
af a CR precursor. The downstream temperature of shocked ions is $6.7 \cdot
10^7$ K, which produces a broad Balmer line with a width of 1774 km/s; in the
absence of acceleration the same width would be 2039 km/s. As already pointed
out, this difference arises because when the acceleration is efficient a non
negligible fraction of the total shock kinetic energy is channelled into CRs
rather than into thermal energy.
% We conclude saying that shocks able to accelerate CRs efficiently wich
% propagate into partially neutral plasma can significantly affects the Balmer
% line emission coming from neutral hydrogen, hence Balmer lines can be used as
% a  diagnostic tools to check the CR acceleration efficiency.

% \section*{Acknowledgments}
% Here the Acknowledgments...

\end{document}